\documentstyle[12pt,epsfig]{article}

\begin{document}
\begin {center}
{\bf {\Large
Forward coherent $\omega$ and $\phi$ mesons production in the photon induced
reaction on nuclei
} }
\end {center}
\begin {center}
Swapan Das \footnote {email: swapand@barc.gov.in} \\
{\it Nuclear Physics Division,
Bhabha Atomic Research Centre,  \\
Trombay, Mumbai: 400085, India \\
Homi Bhabha National Institute, Anushakti Nagar,
Mumbai-400094, India }
\end {center}

\begin {abstract}

The differential cross sections of the forward coherent $\omega$ and $\phi$
mesons photoproduction from nuclei have been calculated using the Glauber
model for the nuclear reaction.
The
measured cross section of the elementary reaction, i.e.,
$\gamma N \to \omega (\phi) N$ reaction, are used as input in the Glauber
model. The experimentally determined free-space scattering parameters of the
quoted mesons have been used to evaluate the meson nucleus interactions.
The
sensitivity of the cross section to the Fermi-motion of the nucleon and the
nucleon-nucleon short-range correlation in the nucleus are studied. The
calculated results are compared with the measured cross sections.

\end {abstract}



\smallskip

\section{Introduction}

The coherent meson production in the nuclear reaction provides opportunities
to learn enriched physics associated the dynamics of the reaction.
In
the resonance region ($\sim$ 1-2 GeV), the meson is originated due to the
decay of the hadronic (or hyperonic) resonance produced in the nuclear
reaction. Therefore, the properties of the resonance in a nucleus can be
investigated by studying the coherent meson production in the nuclear
reaction.
The
coherent pion produced in the $\Delta (1232)$-excitation region had been
extensively used to study the $\Delta$-dynamics in the nucleus \cite{eric,
dasC, korf, chiba}; as the branching ratio of this resonance decaying to
the $N\pi$ channel is $99.4\%$ \cite{tana}.
The
$\Delta$-peak shift in the inclusive nuclear reaction (e.g.,
$(^3\mbox{He},t)$ reaction on the nucleus) relative to the $pp$ collision
\cite{gaar} has been understood due to the coherent $\pi^+$ meson production
in the reaction \cite{chiba}.
The
coherent pion cannot be produced in the $pp$ reaction.
C$\acute{\mbox{o}}$rdoba et al., \cite{dasC} have emphasized that the
$\pi^0$ meson coherently produced in the $(p,p^\prime)$ reaction on a nucleus
can be used as beam.
The
coherent pion production (arising due to $\Delta$-decay) parallel to the
momentum transfer to the nucleus is suppressed in the photo- and
electro-induced nuclear reactions, unlike that occurred in the hadron induced
nuclear reaction \cite{dasC, korf}.
The
$\Delta$-renormalization in the nucleus leads to the reduction of the cross
section of the coherent double pion $\pi\pi$ photoproduction from the nucleus,
and that also modifies the shape of the pion energy distribution
spectrum \cite{kama}.

The coherent $\eta$ meson production from the nucleus has been used to
investigate the dynamics of the resonances heavier than $\Delta (1232)$.
The
$N(1520)$ resonance (though possessing negligibly small branching ratio
$\sim 0.08 \%$ in the $N\eta$ channel \cite{tana}) contributes dominantly to
the coherent $\eta$ meson photoproduction from the spin-isospin saturated
nucleus \cite{benn}.
It
should be mentioned that the resonance $N(1535)$ has large branching ratio
$\sim 50 \%$ in the $N\eta$ channel \cite{tana}, but that is suppressed in
the quoted reaction because the $\gamma NN(1535)$ coupling constants for the
proton and neutron in the nucleus substantially cancel each other
\cite{pet, fix}.
The
consideration of the non-local effect leads to the sizeable contribution of
the $N(1535)$ resonance in the stated reaction \cite{pet}. The importance of
the $t$-channel $\omega$ meson-exchange interaction in the above reaction is
elucidated in Refs.~\cite{pet, fix}.
The
contributions of the hadronic resonances to the coherently produced $\eta$
meson in the proton nucleus collision have been explored \cite{dasE}, which
reveal the dominant contribution to the coherent $\eta$ meson production
arises due to the decay of $N(1520)$, specifically, at high energy.
In
search of $\eta$-mesic nucleus, the importance of $N(1535)$ resonance is
mentioned in the coherent $\eta$ meson photoproduction from $^3$He nucleus
\cite{pher}.

Beyond the resonance region, the coherent photoproductions of the $\omega$
and $\phi$ mesons from the deuteron at high momentum transfer region could
not be explained by considering only the single scattering of the meson but
those are realized by including the double scattering of the meson
\cite{chet, fran}.
The
semi-hadronic decays of the $\rho^0$ and $\omega$ mesons coherently
photoproduced in the nucleus demonstrates the enhancement in the cross
section away from the peak region due to the interference of these mesons
\cite{dasS}.
The
coherent $\rho$ meson photoproduction from the nucleus was studied to extract
the informations about the $\rho-N$ coupling constant, $\rho-N$ scattering
amplitude and $N-N$ correlation in the nucleus \cite{gott}.
The
modification of the hadronic parameters (i.e., mass and width) is looked for
the $\rho$ and $\omega$ mesons coherently produced in the photonuclear
reactions \cite{dasM}.

The nuclear effect on the vector meson scattering parameters (i.e., the ratio
of the real to imaginary part of the vector meson nucleon scattering
amplitude $\alpha^*_{VN}$ and the vector meson nucleon scattering cross
section $\sigma^{*VN}_t$ in the nucleus) is investigated by studying the
nuclear mass number $A$ dependent forward coherent four-momentum transfer
cross section $\frac{d\sigma^{\gamma V}_{coh}}{dq^2} (0)$ of the
$\gamma A \to VA$ reaction. The symbol $V$ is used to represent a vector
meson, e.g., either $\omega$ or $\phi$ meson.
Using
the vector meson dominance (VMD) model for the production of the quoted
meson in the Glauber model, Sibirtsev et al., \cite{sibi, sibi1} have
analyzed the measured $\frac{d\sigma^{\gamma V}_{coh}}{dq^2} (0)$
\cite{brod, mccl} to extract $\alpha^*_{VN}$ and $\sigma^{*VN}_t$ for the
$\omega$ and $\phi$ mesons. The Fermi-motion and short-range correlation of
the nucleons in the nucleus are not considered in this analysis.

The quoted data for $\frac{d\sigma^{\gamma V}_{coh}}{dq^2} (0)$
\cite{brod, mccl} have been reanalyzed using Glauber model where the
production of the $\omega$ and $\phi$ mesons has been described by the
respective measured forward four-momentum transfer cross section (i.e.,
$\frac{d\sigma}{dq^2} ^ {\gamma N \to \omega N}$ and
$\frac{d\sigma}{dq^2}^{\gamma N \to \phi N}$) of the photonucleon reaction.
The
experimentally determined free-space scattering parameters (i.e.,
$\alpha_{VN}$ and $\sigma^{VN}_t$) for the above mesons are used to evaluate
the distortion arising due to the meson nucleus interaction.
The
sensitivity of the calculated cross section
$\frac{d\sigma^{\gamma V}_{coh}}{dq^2} (0)$ to the Fermi-motion and
short-range correlation of the nucleons in the nucleus has been investigated.
The
formalism of the reaction has been presented in sec.~2. The calculated
results are discussed and compared with the data in sec.~3. The last section,
i.e., sec.~4, ends with the conclusion.

\section{Formalism}

The differential cross section $\frac{d\sigma^{\gamma V}_{coh}}{dq^2}$ of
the four-momentum transfer distribution for the coherent vector meson
production in the photonuclear reaction, according to Glauber model
\cite{frank, bauer}, is given by
\begin{equation}
\frac{d\sigma^{\gamma V}_{coh}}{dq^2}
=
\frac{d\sigma}{dq^2} ^ {\gamma N \to VN} 
\left | \int ... \int d{\bf b} dz \varrho (b,z) 
 e^{ i( {\bf q}_\perp \cdot {\bf b} + q_\parallel z )  }
 D^{(-)*}_{k_V} (b,z) \right |^2,
\label{gAA}
\end{equation}
where $V$ denotes either $\omega$ or $\phi$, as mentioned earlier. $q^2$ is
the four-momentum transfer to the nucleus. $q_\parallel$ and ${\bf q}_\perp$
are the parallel and perpendicular components of the momentum transfer vector.
$\varrho (b,z)$ is the density distribution of the nucleus, normalized to the
mass number of the nucleus.

$\frac{d\sigma}{dq^2}^{\gamma N \to VN}$ in the above equation represents
the cross section for the four-momentum transfer distribution in the
elementary $\gamma N \to VN$ reaction.
The
Glauber model is based on the fixed scatterer or frozen nucleon approximation
\cite{glaub}. Therefore, the Fermi-motion of the nucleon in the nucleus has
been incorporated replacing $\frac{d\sigma}{dq^2}^{\gamma N \to VN}$ by that
in the nucleus, i.e.,
$\left < \frac{d\sigma}{dq^2}^{\gamma N \to VN} \right >_A$.
For the beam energy $E_\gamma$ and forward emission of the vector meson,
it can be written as
\begin{equation}
\left <
\frac{d\sigma}{dq^2} ^ {\gamma N \to VN} (0, E_\gamma)
\right >_A
= \int \int d{\bf k}_N d\epsilon_N S_A ({\bf k}_N, \epsilon_N) 
  \frac{d\sigma}{dq^2} ^ {\gamma N \to VN} (0, E^\prime_\gamma),
\label{gNN}
\end{equation}
with
\begin{eqnarray}
E^\prime_\gamma &=& \frac{s-m^{*2}_N}{2m^*_N}; ~~~
                                    m^*_N=m_N-\epsilon_N,  \nonumber   \\
s &=& (E_\gamma+E_N)^2 - ({\bf k}_\gamma + {\bf k}_N)^2,    \nonumber  \\
E_N &=& m_A - \sqrt{ (k_N)^2+(m_A - m^*_N)^2 }.   \nonumber
\label{sEN}
\end{eqnarray}
The spectral function $S_A ({\bf k}_N, \epsilon_N)$ of the nucleus,
normalized to unity, describes the probability of a nucleon with momentum
${\bf k}_N$ and binding energy $\epsilon_N$ in the nucleus \cite{pary}.
$S_A ({\bf k}_N, \epsilon_N)$ for various nuclei are discussed elaborately
in Ref.~\cite{pary2}. Therefore, those have not been presented explicitly.

$D^{(-)*}_{k_V} (b,z)$ in Eq.~($\ref{gAA}$) describes the distortion arising
due to the interaction of the vector meson with the nucleus. According to
the eikonal description in the Glauber model \cite{dasS, glaub}, it is given
by
\begin{equation}
D^{(-)*}_{k_V} (b,z) =
exp \left [-\frac{i}{v_V} \int^\infty_z dz^\prime V_{OV}(b,z^\prime) \right ],
\label{DfA}
\end{equation}
where
$V_{OV}$ denotes the optical potential due to the meson nucleus interaction.
It can be expressed as
\begin{equation}
\frac{1}{v_V}V_{OV} (b,z^\prime) 
=-\frac{1}{2} (\alpha_{VN}+i) \sigma^{VN}_t \varrho (b, z^\prime),
\label{Opf}
\end{equation}
with
$\alpha_{VN}$ representing the ratio of the real to imaginary part of the
forward scattering amplitude $f_{VN \to VN} (0)$ of the elementary
$VN \to VN$ reaction in the free-space. $\sigma^{VN}_t$ denotes the total
vector meson nucleon scattering cross section:
$\sigma^{VN}_t$ = $\frac{4\pi}{k_V} f_{VN \to VN} (0)$.

It should be mentioned that the nucleons bound in the nucleus do not occupy
the same spatial region (called nuclear granularity \cite{lee}) because of
the nucleon-nucleon short-range correlation which arises due to the repulsive
(short-range) interaction between the nucleons in the nucleus.
This
correlation prevents the shadowing of the vector meson nucleon interaction
due to the surrounding nucleons present in the nucleus. To include
the nucleon-nucleon short-range correlation, $\varrho (b,z^\prime)$ in
Eq.~($\ref{Opf}$) must be replaced \cite{lee} as
\begin{equation}
\varrho (b,z^\prime) \to \varrho (b,z^\prime) C(l),
\label{src}
\end{equation}
where $C(l)$ representing the correlation function depends on the path-length
$l (=|z^\prime -z|)$ traversed by the vector meson. The nuclear matter
estimation for $C(l)$ \cite{lee, mill} is
\begin{equation}
C(l) = \left [ 1-\frac{h(l)^2}{4}  \right ]^{1/2} [1+f(l)],
\label{ppl}
\end{equation}
with
\begin{equation}
h(l) = 3\frac{j_1 (k_F l)}{k_F l} ~ \mbox{and} ~
f(l) =-e^{-\alpha l^2} (1-\beta l^2).
\label{ppl2}
\end{equation}
The values of the parameters $\alpha$ and $\beta$ are 1.1 and 0.68 fm$^{-2}$
respectively. The Fermi momentum is taken equal to 1.36 fm$^{-1}$. This
simplified version of the correlation function agrees well with those
derived using detail many-body calculations \cite{lee}.

\section{Result and Discussions}

The differential cross sections $\frac{d\sigma^{\gamma V}_{coh}}{dq^2}(0)$
of the forward coherent vector mesons, i.e., $\omega$ and $\phi$ mesons,
photoproduction from nuclei have been calculated using Glauber model in the
multi-GeV region.
The
cross sections are also evaluated considering the Fermi-motion of the nucleon
and the nucleon-nucleon short-range correlation in the Glauber model. 
The
energy dependent measured cross section for the forward vector meson
production in the $\gamma N \to VN$ reaction, i.e.,
$\frac{d\sigma}{dq^2}^{\gamma N \to VN}$ used in Eqs.~(\ref{gAA}) and
(\ref{gNN}), are taken from Refs.~\cite{sibi1, sibi3, mibe}.
The
density distribution $\varrho (r)$ of deuteron is generated using its
wave-function due to Paris potential \cite{lacom}. $\varrho (r)$ for $^{12}$C
nucleus is described by the harmonic oscillator Gaussian form, where as
that for other nuclei (heavier than $^{12}$C) is illustrated by the
two-parameter Fermi distribution function \cite{andt}.

The optical potential $V_{OV}$ in Eq.~(\ref{Opf}) is evaluated using the
free-space vector meson nucleon scattering parameters $\alpha_{VN}$ and
$\sigma^{VN}_t$.
For the $\omega$ meson, $\alpha_{\omega N}$ is taken from the calculation
of the additive quark model and Regge theory due to Donnachie and Landshoff
\cite{sibi} (see the references there in). The experimentally determined
values of $\sigma^{\omega N}_t$ are given in Refs.~\cite{sibi,lyka}.
Using
the vector meson dominance model,
$ \frac{d\sigma}{dq^2}^{\gamma N \to \phi N}(0) $ can be written
\cite{sibi1, das96} as
\begin{equation}
\frac{d\sigma}{dq^2} ^ {\gamma N \to \phi N} (0)
= \frac{\alpha_{em}}{16\gamma^2_\phi}
  \left ( \frac{ \tilde{k}_\phi }{ \tilde{k}_\gamma } \right )^2
   [1+\alpha^2_{\phi N}] (\sigma^{\phi N}_t)^2,
\label{gNN2}
\end{equation}
where $\tilde{k}_\gamma$ and $\tilde{k}_\phi$ are the momenta in the
$\gamma N$ and $\phi N$ center of mass systems respectively, evaluated at
the energy in the $\gamma N$ center of mass system.
$\alpha_{em} (=1/137.036)$
is the fine structure constant. $\gamma_\phi (=6.72)$ is related to the
photon coupling to $\phi$ meson \cite{saku}, which is extracted from
the measured $\phi \to e^+e^-$ decay width \cite{tana} using the expression
$\Gamma (m_\phi)_{\phi \to e^+e^-}$ = $\frac{\pi}{3}
(\frac{\alpha_{em}}{\gamma_\phi})^2 m_\phi$ \cite{saku}. 
The
energy dependent values of $\sigma^{\phi N}_t$ have been evaluated using
the experimentally determined values of
$\frac{d\sigma}{dq^2} ^ {\gamma p \to \phi p} (0)$ \cite{sibi1, mibe} and
$\alpha_{\phi N} =-0.3$  \cite{bauer} in the above equation.

The mass number $A$ dependent cross sections
$(1/A)$$d\sigma^{\gamma \omega}_{coh}/dq^2 (0)$ of the forward coherent
$\omega$ meson production in the photonuclear reactions, calculated at 3.9
GeV, are shown in Fig.~\ref{FgOC}.
The
short-dashed curve (labeled as GM) represents the calculated results due to
Glauber model. The large-dashed curve (labeled as GM+FM) illustrates that
due to the Fermi-motion of the nucleon considered in the Glauber model.
This
figure shows the decrease in the calculated cross section due to the
Fermi-motion. It occurs because the measured forward
$ \frac{d\sigma}{dq^2}^{\gamma N \to \omega N} (0,E_\gamma) $, as shown in
Refs.~\cite{sibi1, sibi3}, is very sensitive to the energy, i.e., the cross
section sharply falls with the increase in the energy around
$E_\gamma=3.9$ GeV.
As
explained in Eq.~(\ref{gNN}), the forward
$\frac{d\sigma}{dq^2}^{\gamma N \to \omega N} (0,E^\prime_\gamma)$ is to be
evaluated at the beam energy $E^\prime_\gamma$ $(\neq E_\gamma)$ while the
Fermi-motion of the nucleon is considered.
The
dot-dashed curve (labeled as GM+SR) denotes the calculated cross section due
to the nucleon-nucleon short-range correlation (without Fermi-motion)
incorporated in the Glauber model. It shows the cross section is drastically
increased, i.e., the absorption of the $\omega$ meson is reduced due to the
short-range correlation of the nucleons in the nucleus.
The
solid curve (labeled as GM+FS) arises because of the inclusion of both
effects (i.e., Fermi-motion and short-range correlation) in the Glauber
model. The dashed and solid curves in the figure elucidate that the cross
section increases for the nuclei $A > 25$.
In
Fig.~\ref{FgOD}, the calculated cross sections (described by the dashed and
solid curves) are compared with the data \cite{brod}.

The variation of the calculated forward coherent $\phi$ meson photoproduction
cross sections, i.e., $(1/A)$$d\sigma^{\gamma \phi}_{coh}/dq^2 (0)$, with
nuclei at 6.4 and 8.3 GeV are presented in Fig.~$\ref{FgPC}$. The
short-dashed curve (labeled as GM) shows the result calculated using Glauber
model.
The
contribution of the Fermi-motion of the nucleon to the cross section (not
shown in the figure) is found insignificant for both energies.
It
occurs since the elementary cross section, i.e.,
$ \frac{d\sigma}{dq^2} ^ {\gamma N \to \phi N} (0,E^\prime_\gamma) $ in
Eq.~(\ref{gNN}), is not sensitive to the beam energy in the above mentioned
energy region \cite{sibi1, mibe}. This is unlike to that occurs for the
$\omega$ meson photoproduction at 3.9 GeV, as discussed earlier.
The
dot-dashed curve (labeled as GM+SR) shows the cross section (similar to
that depicted for the $\omega$ meson) increases due to the short-range
correlation of the nucleons in the nucleus.

The cross section $(1/A)$$d\sigma^{\gamma \phi}_{coh}/dq^2 (0)$ calculated
using the Glauber model (dashed curve) and that calculated considering the
Fermi-motion and short-range correlation of the nucleons in the Glauber model
(solid curve) have been presented along with the data \cite{mccl} in
Fig.~\ref{FgPD}.
As
mentioned earlier, the contribution of the Fermi-motion to the cross section
is negligibly small. Therefore, the cross sections measured at 6.4 and 8.3
GeV (as shown in the figure) is better understood (especially for heavy
nuclei) because of the nucleon-nucleon short-range correlation incorporated
in the Glauber model.

\section{Conclusions}

The forward coherent vector $(V)$ mesons, i.e., $\omega$ and $\phi$ mesons,
production cross sections $\frac{d\sigma^{\gamma V}_{coh}}{dq^2}(0)$ in the
photonuclear reactions have been calculated in the multi-GeV region using
Glauber model,
where
the measured forward cross section $\frac{d\sigma}{dq^2}^{\gamma N \to VN}$
of the $\gamma N \to VN$ reaction and the experimentally determined
free-space scattering parameters $\alpha_{VN}$ and $\sigma^{VN}_t$ are used.
The
sensitivity of the cross section to the Fermi-motion and nucleon-nucleon
short-range correlation of the nucleons in the nucleus is investigated.
The
calculated results for the $\omega$ meson photoproduction from nuclei at 3.9
GeV show that the cross section is sensitive to both of these effects. The
cross section decreases due to the Fermi-motion where as that increases
because of the nucleon-nucleon short-range correlation.
The
cross sections of the $\phi$ meson photoproduction from nuclei at 6.4 and
8.3 GeV are found insensitive to the Fermi-motion, but those are considerably
enhanced due to the short-range correlation.
The
cross sections $\frac{d\sigma^{\gamma V}_{coh}}{dq^2}(0)$ calculated using
the measured $\frac{d\sigma}{dq^2}^{\gamma N \to VN}$, $\alpha_{VN}$ and
$\sigma^{VN}_t$ for both $\omega$ and $\phi$ mesons are well accord with the
data.

\section{Acknowledgement}

The author duly acknowledges Dr. S. M. Yusuf for his support to carry out
this work.

\newpage

\begin{figure}[h]
\begin{center}
\centerline {\vbox {
\psfig{figure=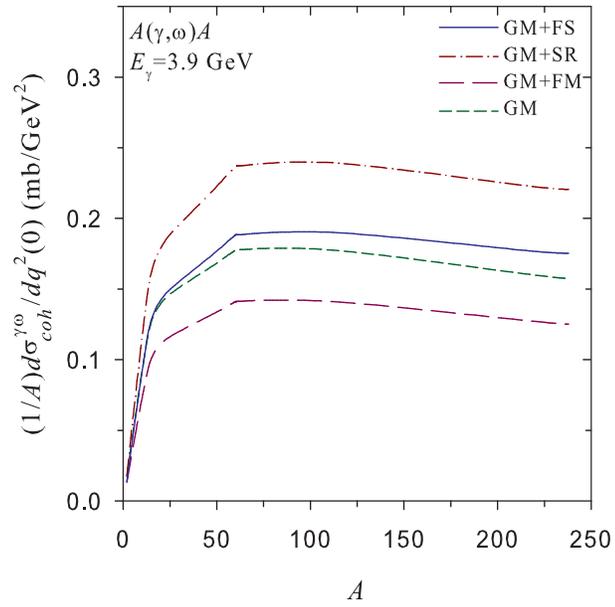,height=08.0 cm,width=08.0 cm}
}}
\caption{
(color online).
The calculated nuclear mass number $A$ dependent cross sections of the
forward coherent $\omega$ meson production in the $\gamma A$ reaction.
Various curves in the figure are elaborated in the text.
}
\label{FgOC}
\end{center}
\end{figure}

\begin{figure}[h]
\begin{center}
\centerline {\vbox {
\psfig{figure=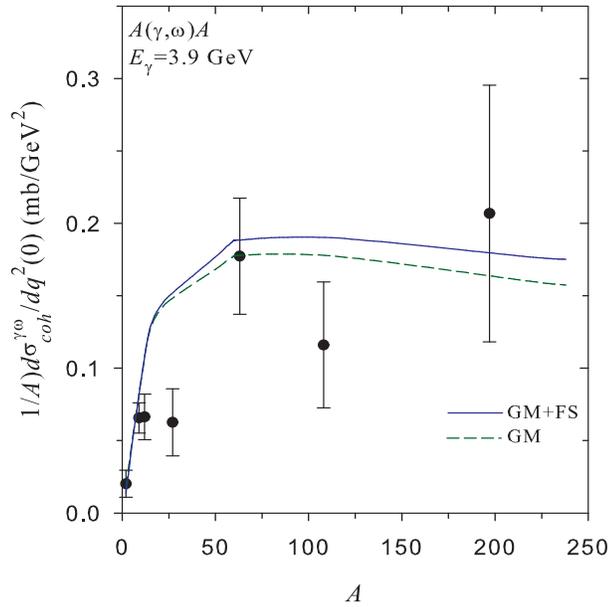,height=08.0 cm,width=08.0 cm}
}}
\caption{
(color online).
The calculated results for the $\omega$ meson are compared with the data
\cite{brod}. The dashed curve describes the cross section calculated using
Glauber model, where as the solid curve arises because of the Fermi-motion
and short-range correlation of the nucleons in the nucleus incorporated in
the Glauber model.
}
\label{FgOD}
\end{center}
\end{figure}

\begin{figure}[h]
\begin{center}
\centerline {\vbox {
\psfig{figure=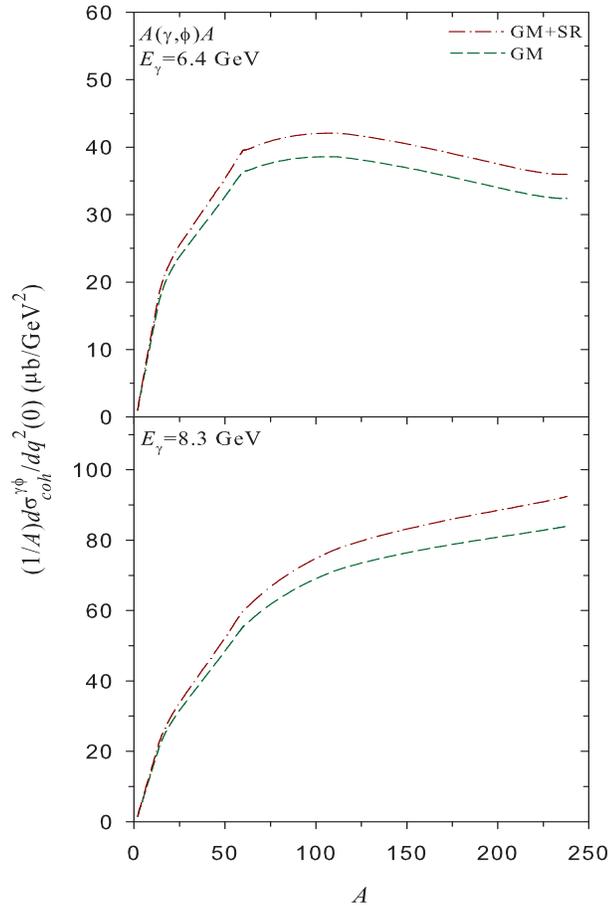,height=12.0 cm,width=08.0 cm}
}}
\caption{
(color online).
Same as those presented in Fig.~\ref{FgOC} but for the $\phi$ meson at
different energies. Since the contribution of the Fermi-motion of the nucleon
to the cross section is negligibly small, it is not shown in the figure.
}
\label{FgPC}
\end{center}
\end{figure}

\begin{figure}[h]
\begin{center}
\centerline {\vbox {
\psfig{figure=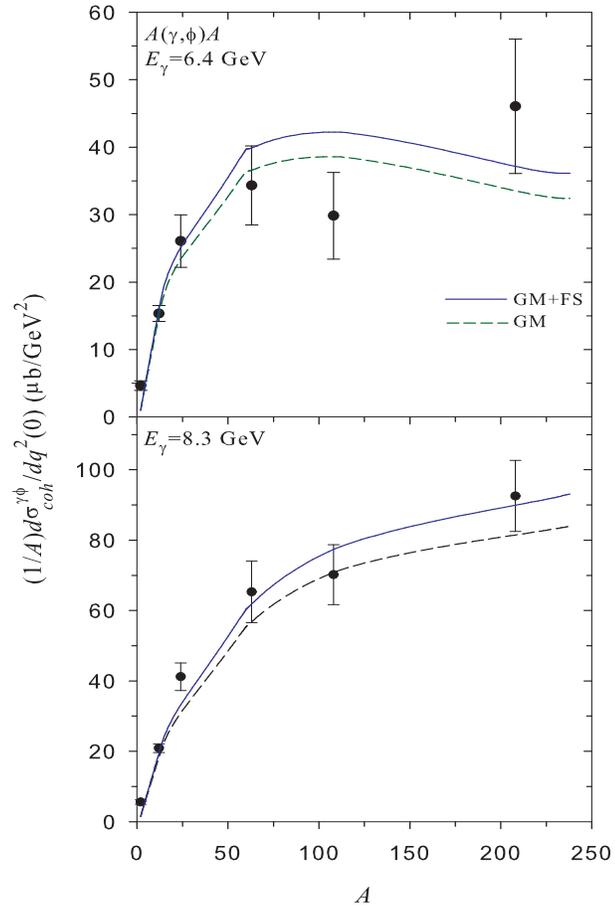,height=12.0 cm,width=08.0 cm}
}}
\caption{
(color online).
Same as those in Fig.~\ref{FgOD} but for the $\phi$ meson at different
energies. The data are taken from Ref.~\cite{mccl}.
}
\label{FgPD}
\end{center}
\end{figure}

{\bf Figure Captions}
\begin{enumerate}

\item
(color online).
The calculated nuclear mass number $A$ dependent cross sections of the
forward coherent $\omega$ meson production in the $\gamma A$ reaction.
Various curves in the figure are elaborated in the text.

\item
(color online).
The calculated results for the $\omega$ meson are compared with the data
\cite{brod}. The dashed curve describes the cross section calculated using
Glauber model, where as the solid curve arises because of the Fermi-motion
and short-range correlation of the nucleons in the nucleus incorporated in
the Glauber model.

\item
(color online).
Same as those presented in Fig.~\ref{FgOC} but for the $\phi$ meson at
different energies. Since the contribution of the Fermi-motion of the nucleon
to the cross section is negligibly small, it is not shown in the figure.

\item
(color online).
Same as those in Fig.~\ref{FgOD} but for the $\phi$ meson at different
energies. The data are taken from Ref.~\cite{mccl}.

\end{enumerate}

\newpage

\newpage

\newpage

\newpage

\end{document}